\documentclass[3p, final, authoryear]{elsarticle}
\usepackage{graphicx}
\usepackage{latexsym}

\usepackage{amsmath}
\usepackage{graphicx}
\usepackage{textcomp}
\usepackage{amstext}
\usepackage{amssymb}
\usepackage{tikz}
\usepackage{lettrine}
\usepackage[export]{adjustbox}
\usepackage{ragged2e}
\usepackage{subfig}
\usepackage{float}
\usepackage{booktabs}
\usepackage{makecell, multirow}
\usepackage{natbib}
\begin{document}

\begin{frontmatter}

\journal{arXiv}

\title{Hedging Properties of Algorithmic Investment Strategies using Long Short-Term Memory and Time Series models for Equity Indices\tnoteref{t2}}

\tnotetext[t2]{This document is the results of the research project funded by IDUB program: BOB-IDUB-622-233/2022 at University of Warsaw}


\author[1,2]{Jakub Michańków\corref{cor1}%
\fnref{fn1}}
\ead{jmichankow@wne.uw.edu.pl}

\author[2]{Paweł Sakowski\fnref{fn2}}
\ead{sakowski@wne.uw.edu.pl}

\author[2]{Robert Ślepaczuk\fnref{fn3}}
\ead{rslepaczuk@wne.uw.edu.pl}

\cortext[cor1]{Corresponding author: jmichankow@wne.uw.edu.pl}

\fntext[fn1]{ORCID: 0000-0002-0567-6240}
\fntext[fn2]{ORCID: 0000-0003-3384-3795}
\fntext[fn3]{ORCID: 0000-0001-5227-2014}

\affiliation[1]{
    organization={Department of Informatics, Krakow University of Economics},
    addressline={ul. Rakowicka 27},
    city={Krakow},
    postcode={31-510},
    country={Poland}
}
\affiliation[2]{
    organization={Quantitative Finance Research Group, Department of Quantitative Finance, Faculty of Economic Sciences, University of Warsaw},
    addressline={ul. Długa 44/50},
    postcode={00-241},
    city={Warsaw},
    country={Poland}
    }

\date{September 2023}

\begin{abstract}
This paper proposes a novel approach to hedging portfolios of risky assets when financial markets are affected by financial turmoils. We introduce a completely novel approach to diversification activity not on the level of single assets but on the level of ensemble algorithmic investment strategies (AIS) built based on the prices of these assets. We employ four types of diverse theoretical models (LSTM - Long Short-Term Memory, ARIMA-GARCH - Autoregressive Integrated Moving Average - Generalized Autoregressive Conditional Heteroskedasticity, momentum, and contrarian) to generate price forecasts, which are then used to produce investment signals in single and complex AIS. In such a way, we are able to verify the diversification potential of different types of investment strategies consisting of various assets (energy commodities, precious metals, cryptocurrencies, or soft commodities) in hedging ensemble AIS built for equity indices (S\&P 500 index). Empirical data used in this study cover the period between 2004 and 2022. Our main conclusion is that LSTM-based strategies outperform the other models and that the best diversifier for the AIS built for the S\&P 500 index is the AIS built for Bitcoin. Finally, we test the LSTM model for a higher frequency of data (1 hour). We conclude that it outperforms the results obtained using daily data.
\end{abstract}

\end{frontmatter}

\section{Introduction}
\label{Introduction}

The main objective of this research is to improve the decision-making process by incorporating energy commodities and other asset classes into the hedging strategy of a diversified portfolio comprised of ensemble algorithmic investment strategies (AIS) constructed for the S\&P 500 index. We present novel multidimensional verification of the possibilities of constructing and combining algorithmic investment strategies developed on the basis of 1) the Long Short-Term Memory (LSTM) model, 2) the ARIMA-GARCH class models, as well as concepts of 3) contrarian and 4) momentum strategies for various assets: equity indices, precious metals, energy and soft commodities, and cryptocurrencies. The selection of theoretical models and assets is dictated by the aim to include a set of those which is diverse enough and at the same time highly tested in the literature. We are going to achieve it by:

- testing the efficiency of single strategy and ensemble strategies built with: 1) various types of assets, 2) various theoretical models,

- introducing a walk-forward approach enabling us to test theoretical models on various training, validation, and testing periods with different characteristics of return distributions,

- verifying the diversification potential of various strategies built using different theoretical concepts and different types of assets in hedging investment strategies built on the S\&P500 index, 

- performing sensitivity analysis in order to check the robustness of final results to various frequencies of data.

Our main contribution to existing literature can be found in a completely novel approach to testing diversification and hedging potential. We focus on the combination of single and ensemble algorithmic investment strategies built for various types of assets in order to maximize risk-adjusted return instead of focusing on just a single combination of new assets with adequate characteristics of returns enabling us to optimize the weights of our portfolio.

In our research, we use a walk-forward procedure on a daily time series with dates ranging from 2004-01-02 to 2022-03-29.  In practice, the starting point of data depends on the asset and availability of data for the tested asset and varies between 2004-01-02 and 2010-07-17. In order to accomplish the main aim we decided to formulate the following research questions (RQ):

- RQ1: \textit{Which of the tested groups of assets (energy commodities, cryptocurrencies, gold, or soft commodities) has the largest diversification potential in complex AIS (built with machine learning (ML) models and ARIMA-GARCH models) for equity indices?}

- RQ2: \textit{Are ML techniques more efficient than ARIMA-GARCH models and the concepts of momentum and contrarian in the case of single and complex (ensemble model combining all tested strategies for the given assets - type I) investment strategies.}

- RQ3: \textit{Are complex (ensemble) AIS based on the aggregation of all theoretical models for the single asset (type I) or all assets for a single theoretical model (type II) more efficient than individual strategies?}

- RQ4: \textit{Are results for LSTM models on higher frequencies of data (1h) better than those on daily data.}

The problem analyzed in this research is a fundamental issue not only from the micro, but also from the macro point of view, especially if we realize how much the stability of the financial systems of individual countries, and the state of savings of their citizens, are affected by the efficient and effective asset management in mutual and pension funds, investment funds, hedge funds or insurance companies. Wrong decisions in the allocation of these assets, especially in the context of long-term investment policies and specific investment strategies in the medium-term have very important consequences in the context of financial security and the quality of life of citizens of these countries. A similar approach to the one presented in this paper could also be extended to financial risk or macroeconomic forecasting.

The structure of this paper is as follows. After the introduction in Section \ref{Introduction}, we present a comprehensive literature review in Section \ref{Literature review}. Then, in Section \ref{Methodology and Data} we describe the details of methodology and data. Finally, Section \ref{Results} covers the main results and Section \ref{Conclusions} presents conclusions.

\section{Literature review}
\label{Literature review}

In this short literature review, we present a historical background covering the development of (recurrent neural networks) RNN and long short-term memory (LSTM) models and the summary of various empirical papers testing the efficiency of LSTM on various types of assets, frequencies, and studies trying to ensemble it in different ways.

\cite{hochreiter_long_1997} are responsible for the introduction of LSTM. By introducing Constant Error Carousel (CEC) units, LSTM deals with the exploding and vanishing gradient problems. The initial version of the LSTM block included cells, input, and output gates. \cite{gers_learning_2000} introduced the forget gate (also called “keep gate”) into LSTM architecture, enabling the LSTM to reset its own state. They added peephole connections (connections from the cell to the gates) into the architecture. Additionally, the output activation function was omitted. Then, \cite{chung_empirical_2014} put forward a simplified variant called Gated Recurrent Unit (GRU).
 
\cite{chen2015lstm} implemented the LSTM model to predict the next-day returns for China stocks. \cite{zhang_at-lstm_2019} presented the AT-LSTM model which is the combination of LSTM and
Attention-based model. They provided results for three index datasets: Russell 2000, DJIA, and NASDAQ, and argued that their framework for time series prediction is state-of-the-art against the baselines. \cite{kijewskia2020predicting} compared the performance of classical techniques with the LSTM model for the S\&P500 index on daily frequency for the last 20 years and showed that LSTM model results are highly dependent on initial hyperparameters assumptions. \cite{siami2018comparison} investigate whether and how newly identified deep learning time series forecasting algorithms, such as LSTM, outperform more seasoned ones. It is discovered that LSTM and other deep learning algorithms outperform more traditional algorithms like the ARIMA model. More specifically, LSTM outperformed ARIMA by achieving an average error rate reduction that was between 84 and 87 percent lower.

\cite{castellano2021slepaczuk} tested the portfolio of algorithmic investment strategies (TA indicators, calendar anomalies, Macro, and ARIMA models) built on S\&P500 and Nasdaq Composite indices in the period of the last 40 years. They revealed that especially ensemble models can beat the benchmark in times of turbulent events as well as during very fast market growth. \cite{di2017recurrent} analyze the performance of three different recurrent neural network models—a basic RNN, the LSTM, and the Gated Recurrent Unit (GRU) — using the price of Google stock. The authors also go over the RNN's hidden dynamics and provide examples. The data clearly show that on a five-day horizon, the LSTM outperformed other versions with a 72 percent accuracy. \cite{grudniewiczslepaczuk2021} applied several Machine Learning algorithms to technical analysis indicators for the WIG20, DAX, S\&P 500, and a few selected CEE indices. The study’s findings reveal that quantitative techniques beat passive strategies in terms of risk-adjusted returns, with the Bayesian Generalized Linear Model and Naive Bayes being the top models for the investigated indices.

Studies additionally make an effort to combine an ensemble or hybrid technique with LSTM. \cite{Hossain2018HybridDL} created a deep learning hybrid model using the well-known architectures: LSTM, and GRU (2018). The authors train a prediction model using the S\&P 500 index time series, which spans over 66 years (1950 to 2016). This method involves passing the input data to the LSTM network, which generates a first-level prediction, and the output of the LSTM layer to the GRU layer, which generates the final prediction. With an MSE of 0.00098 in prediction, the proposed network outperforms earlier neural network methodologies. \cite{michankow2022etal} compared the use of the LSTM model in AIS on BTC and S\&P500 index on various frequencies. They showed that the efficiency of LSTM in AIS strictly depends on HT and the construction of the model and estimation process. Additionally, they introduced and revealed that proper Loss Function (Mean Absolute Directional Loss - MADL) is crucial in the model estimation process and that the results are dependent on asset classes tested and frequencies used. Their final results were not robust. 
\cite{shah2018comparative} provide a good example of how an LSTM-RNN model may deliver exceptional predictions on non-stationary data (2018). They show that the LSTM model not only yields great outcomes for daily forecasts, or predictions made one day in advance but also yields results that are more than satisfactory for predictions made seven days in advance using only the daily price as a feature. 

\cite{vo2022slepaczuk} compared the performance of ARIMA with the combination of ARIMA and GARCH family models to forecast S\&P 500 index log returns in order to construct algorithmic investment strategies on this index. Their main contribution was that the hybrid models outperformed ARIMA and the benchmark (Buy\&Hold strategy on S\&P 500 index) over the long term. These results were not sensitive to varying window sizes, the type of distribution, and the type of the GARCH model. The current advancements in high-frequency data estimation are related not only to technological issues and the growing processing power of big data but also to the requirement to understand and predict the behavior of variables over shorter time horizons. In their study on high-frequency Bitcoin trading, \cite{lahmiri2020bekiros} used three different kinds of machine learning (ML) models: (i) algorithmic models like regression trees, (ii) statistical ML techniques like support vector regressions (SVR), and (iii) ANN topologies like feedforward (FFNN) or Bayesian regularization (BRNN). Their findings show that artificial neural networks perform better than other types of systems in noisy signal environments. \cite{baranochnikov2022slepaczuk} presented a walk-forward procedure that is in charge of training models and choosing the best one in order to predict future values of financial assets. They test the algorithms on four financial assets (Bitcoin, Tesla, Brent Oil, and Gold) and discover that LSTM outperforms GRU in the vast majority of cases. In order to compare the performance of random forests and LSTM networks (more specifically, CuDNNLSTM) in predicting the directional movements of the stocks that make up the S\&P 500 index out-of-sample from January 1993 to December 2018 for intraday trading, \cite{ghosh2022neufeldSahoo} used both training methodologies. In addition to returns relative to closing prices, they also introduced returns relative to opening prices and intraday returns in their multi-feature setting. In the end, they performed better than the benchmark.

\cite{FLORI2021772} used LSTM signals to improve portfolio performances of pairs trading strategies and showed that LSTM signals contain information that goes above and beyond traditional indicators. Moreover, what is important in our study they revealed that LSTM signals allow for the disentangling of the reversal effect from the momentum effect. Another paper that applied long short-term memory networks to financial market predictions was written by \cite{FISCHER2018654}. LSTM was benchmarked against deep nets, random forests, and logistic regression. It occurred that Long short-term memory networks exhibit the highest predictional accuracy and returns.

Based on this literature review we can conclude that implementation of the forecasts from LSTM models in buy/sell signals can increase the efficiency of investment strategies. Moreover, we observe a growing number of publications on various types of ensemble models that combine frequencies or assets on the level of the given theoretical models or try to develop new investment techniques by joining many kinds of theoretical models in the process of price forecasting. Finally, we can notice that the type of input variables, the type of normalization, and specifically the architecture of the selected ML model can significantly affect the final results.

\section{Methodology and Data}
\label{Methodology and Data}

\subsection{Terminology and Metrics}
\label{TerminologyAndMetrics} 

\vspace{3pt}

The investment strategies we use in this work are based on the forecasts obtained from 1) ARIMA-GARCH class models, 2) the Long Short-Term Memory network (LSTM), the concepts of 3) contrarian, and 4) momentum effects. In the case of ARIMA-GARCH models, we apply the concise rolling walk-forward procedure with various Information Criteria (Akaike Information Criterion - AIC, Bayesian Information Criterion - BIC, Hannan–Quinn Information Criterion - HQC, etc.). For the purpose of LSTM modeling, a custom loss function (MADL) was created as the network performance metric and is used during the training process (\cite{michankow2022etal}). Buy and sell signals that we use for single investment strategies are based on 1-period ahead forecasts of daily returns. 
Strategy performance metrics (aRC, ASD, MDD, IR, IR*, IR**, nObs, nTrades) are calculated using the equity line constructed for each algorithmic investment strategy separately.

The results for the LSTM model were obtained using R (4.1.0) and Python (3.7.10) programming languages. Deep learning libraries used for design, training, and testing the network are Keras 2.5.0 and TensorFlow 2.7.0. The rest of the calculations, as well as graphs and tables, were done using R and RStudio environment. Computer specifications are as follows: AMD Ryzen 7 3700X 3,6GHz, 16GB RAM, NVIDIA GeForce RTX 2060 Super with 270 tensor cores. One full training (number of iterations $\times$ 300 epochs) lasted around 30 minutes for daily data and around 4-8 hours for hourly data.

\subsection{ARIMA-GARCH model}
\label{ARIMAGARCHmodel} 

\vspace{3pt}

We use the combination of ARIMA($p$,$d$,$q$) and GARCH($r$,$s$) models (\cite{tsay_analysis_2010}). The ARIMA-GARCH model can be regarded as an extension of the ARMA model which is the combination of the autoregressive AR($p$) and moving average MA($q$) models for stationary time series. 

The ARIMA($p$,$d$,$q$) process can be written as:
\begin{equation}
\Biggl (1- \sum_{i=1}^{p}\phi_iL^i \Biggl)(1-L)^dy_t
 = c + \Biggl (1- \sum_{j=1}^{q}\theta_jL^j \Biggl)\varepsilon_t
\end{equation}
where: 

$p$ - is the order of autoregressive terms (AR),

$\phi_i$ - are coefficients of the autoregressive terms,

$L$ - is the lag operator, which produces the previous element of the series, eg. $Ly_t = y_{t-1}$,   

$d$ - is the integration order of $y_t$,

$c$ - is the constant term,

$q$ - is the order of moving-average terms (MA),

$\theta_j$ - are the coefficients of the moving-average terms,

$\varepsilon_t$ - is the IID error term.

In this study, log-returns of assets are described by the ARIMA($p$,0,$q$)-GARCH(1,1) model which is given by:

\begin{equation}
r_t = \mu + \sum_{i = 1}^{p}\phi_ir_{t-i} + \sum_{j = 1}^{q}\theta_j\varepsilon_{t-j} + \varepsilon_t
\end{equation}

\begin{equation}
\varepsilon_t = \sqrt{h_t}z_t, \quad \quad z_t \overset{\mathrm{IID}}{\sim}  N(0,1) 
\end{equation}

\begin{equation}
h_t = \omega + \alpha\varepsilon^2_{t-1} + \beta h_{t-1}
\end{equation}
where $\mu$, $\omega$, $\alpha$, $\beta$ are parameters, $z_t$ is the IID error term, and ${h_t}$ is the conditional variance function.

We use the following estimation process in order to prepare forecasts based on ARIMA($p$,0,$q$)-GARCH(1,1) model: the parameters of the model are re-estimated every day; ARMA(p,q) orders are re-optimized every quarter with AIC, SBC, and HQC ($p_{\mathsf{max}}$ = 5, $q_{\mathsf{max}}$ = 5); AIC is used for the base case scenario; when the estimation of the model was not possible we use the last available model.

\vspace{-10pt}

\subsection{Contrarian and momentum strategies}
\label{ContrarianAndMomentum} 

\subsubsection{Contrarian approach}
\label{Contrarian}

\vspace{3pt}

It is one of the simplest investment strategies (\cite{park2011herding}, \cite{dobrynskaya2019avoiding}, \cite{KADOYA2008120}, and \cite{CARTA2022117763}) assuming a strong mean-reverting process in the analyzed time series, which implies that our next day return forecast is exactly opposite to the previous day's return:

\vspace{-5pt}
\begin{equation}
\mathrm{Buy}_{\mathrm{signal}} \quad \mathrm{on} \quad P_t \quad \mathrm{if} \quad r_{t} < 0
\end{equation}
\begin{equation}
\mathrm{Sell}_{\mathrm{signal}} \quad \mathrm{on} \quad P_t \quad  \mathrm{if} \quad r_{t} \geq 0
\end{equation}

where $P_t$ is the price at time $t$.

\subsubsection{Momentum approach}
\label{Momentum}

Momentum strategy (\cite{jegadeesh2011momentum}, \cite{chu2020high}, \cite{FLORI2021772}, \cite{ONG2023120587}, and \cite{PAL2023110457}) assumes that financial returns tend to be persistent, which implies that our next-day return forecast is exactly the same with regard to the sign to the previous day's return:

\begin{equation}
\mathrm{Buy}_{\mathrm{signal}} \quad \mathrm{on} \quad P_t \quad  \mathrm{if} \quad r_{t} \geq 0
\end{equation}
\begin{equation}
\mathrm{Sell}_{\mathrm{signal}} \quad \mathrm{on} \quad P_t \quad  \mathrm{if} \quad r_{t} < 0
\end{equation}
 
In the case of contrarian and momentum signals, their values are based on the return from the previous day.

\vspace{-10pt}

\subsection{LSTM model}
\label{LSTM} 

\subsubsection{Architecture of LSTM}
\label{ArchitectureLSTM}

\vspace{3pt}

\begin{figure}[H]
\centering
\includegraphics[width=\columnwidth]{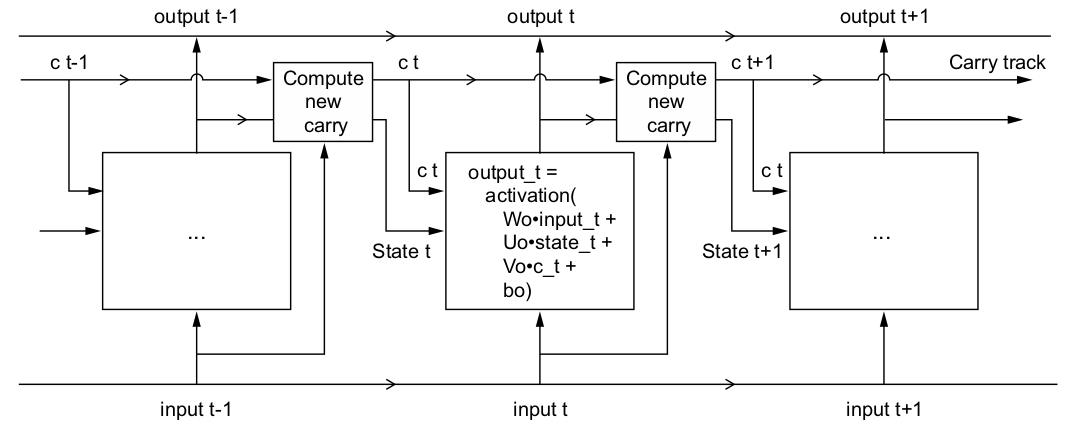}
\scriptsize
\justifying
Note: LSTM cells presented in this figure show the information flow between the main LSTM gates: input, output, and forget. Source: \cite{chollet_2021}
\caption{Anatomy of the LSTM model}
\label{fig:lstm_anatomy}
\end{figure}

LSTM networks (Figure \ref{fig:lstm_anatomy}) are a type of recurrent neural networks (RNNs) that can keep track of long-term dependencies in data, allowing for partial solving of vanishing gradient problems typical for classic RNNs. It's widely used to model sequential data such as text, speech, and time series. LSTM units are composed of memory cells, with each cell having three types of gates (input gate, output gate and forget gate). These gates use "tanh" and "sigmoid" functions to regulate the flow of information through the cell, deciding how much and which information should be stored in a long-term state, passed on to another step, or discarded.

The LSTM adds a way to carry information ($c_t$) across many timesteps and hence preventing older signals from gradually vanishing during processing. The information $c_t$ is combined with the input connection and the recurrent connection:

\begin{equation}
\mathrm{output}_t = f(\mathrm{state}_t \bullet U_o + \mathrm{input}_t \bullet W_o + c_t \bullet V_o + b_o)
\end{equation}

The new value of $c_{t+1}$ is then calculated as:
\begin{equation}
c_{t+1} = i_t \cdot k_t + c_t \cdot f_t
\end{equation}
where:
\begin{equation}
i_t = f(state_t \bullet U_i + input_t \bullet W_i + b_i)
\end{equation}
\begin{equation}f_t = f(state_t \bullet U_f + input_t \bullet W_f + b_f)
\end{equation}
\begin{equation}
k_t = f(state_t \bullet U_k + input_t \bullet W_k + b_k)
\end{equation}
and $U_i$, $W_i$, $b_i$, $U_f$, $W_f$, $b_f$, $U_k$, $W_k$, $b_k$ are matrices with weights,  $f(\cdot)$ is the activation function and $\bullet$ is the dot product of two vectors.

Our LSTM model consists of three LSTM layers with 512/256/128 neurons and one single neuron dense layer on the output. Each of the LSTM layers is using \textit{tanh} activation function (to retain negative values). L2 regularization (0.000001) and dropout (0.001) are also applied to each of these layers. The first two layers return sequences with the same shape as the input sequence (full sequence), and the last LSTM layer returns only the last output.

To train the model we use the Adam optimizer - a stochastic gradient descent optimizer with momentum (estimating first-order and second-order moments). The learning rate of the optimizer is set to 0.5 (after tuning).

\subsubsection{Data selection, hyperparameters tuning and LSTM training}
\label{FEHTaLt}

\vspace{3pt}

We focus primarily on logarithmic returns, using daily data for S\&P500, bitcoin (BTC), gold (GLD), natural gas (UNG), and wheat (ZWF) from 2004-01-02\footnote{In practice, the starting point of data depends on the asset and varies between 2004-01-02 and 2010-07-17.} and 2022-03-29. We also use hourly data for SPX and UNG, from the same period. Hourly data availability is restricted for extensive time periods, so proprietary data was used in this case. However, daily data for all tested assets is readily accessible. 

For the training set, we use an expanding window approach, with the size of the first window set to 252 trading days (one year). The size of the validation set is 33\% of the training set. The test set size is always 252 days. The input sequence size for the LSTM network is set to 10. We use the ReLU activation function on the last neuron to obtain only zero or positive values (for Long Only strategies) or inverted ReLU to obtain zero or negative values (for Short Only strategy used for UNG). The output of the model is a single number predicting the next return value. Based on the sign of the predicted return value we assign -1, 0, and 1 signals, depending on the strategy.

During our research, we conduct detailed hyperparameter tuning to ensure the best possible results from our model. The hyperparameters we test are:

- number of layers (1-5) and neurons in each layer (5-512),

- dropout rate (0 - 0.5) and l2 kernel regularization (0 - 0.01),

- the type of optimizer (SGD, RMSProp, and Adam variants), 

- learning rate (0.0001 - 0.9) and momentum values (0.1-0.9),

- training and testing window sizes, sequence length, and batch size,

- number of epochs (10-300) and callbacks (early stopping and model checkpoint).

\begin{table}[H]
\caption{Values of hyperparameters selected after network tuning.}
\centering
\begin{center}
\begin{tabular}[t]{ll}
\hline
  \textbf{Hyperparameter} & \textbf{Selected Value}\\
\hline
 No. hidden layers & 3\\
 No neurons & 512/256/128\\
 Activation function & \textit{tanh}\\
 Dropout rate & 0\\
 l2 regularizer & 1e-6\\
 Optimizer & Adam\\
 Learning rate & 0.5\\
 Train/test size & 252-exp. window/252\\
 Batch size & exp. window\\
 Sequence length & 10\\
\hline
\end{tabular}
\end{center}
\scriptsize
\justifying
Note: Hyperparameters used in this study for the LSTM model.
\label{tab:Hyperparameters}
\end{table}

In addition, we change the following hyperparameters of the network to optimize it for high-frequency data: train and test sizes are increased to cover one calendar year of data, an additional layer with 252 neurons is added and the number of epochs is changed to 120.

For training and prediction, we use a walk-forward validation/expanding window approach. In the first iteration, the model is trained on one year of data (equal to the train set length) and then used for predictions over the next year (equal to the test set length). After that, the window is expanded by another year of data and the model was retrained. A single return value is predicted each time, using the last 10 (sequence length) values. A single iteration is trained for 300 epochs. The model checkpoint callback function is used to store the best weights (parameters) of the model based on the lowest loss function value in a specific epoch. The weights are then used for prediction.

\subsubsection{Loss function for LSTM model}
\label{LffLm}

\vspace{5pt}

We use the loss function proposed by \cite{michankow2022etal}, who appropriately evaluates the usefulness of the forecasting ability of the LSTM model in algorithmic investment strategies (AIS). The RMSE, MSE, MAE, MAPE, and \%OP used in 99.9\% of similar research are not the proper error functions for the evaluation of the forecasting ability of the given model in the AIS, mainly because they evaluate the point forecast. These error metrics evaluate only the accuracy of forecasts, which is often confused with the forecasting ability of the given model in AIS.

\vspace{-10pt}
\begin{equation}
\mathrm{MADL} =\frac{1}{N} \sum_{i=1}^{N} (-1) \times \mathrm{sign}({R_{i} \times \hat{R}_{i}}) \times \mathrm{abs}(R_{i} )
\end{equation}
\vspace{-10pt}

where:

- $\mathrm{MADL}$ - the Mean Absolute Directional Loss function, 

- \(R_{i}\) is the observed return on interval \(i\), 

- \(\hat{R}_{i}\) is the predicted return on interval \(i\), 

- $\textrm{sign}(X)$ is the function which gives the sign of \(X\),

- $\textrm{abs}(X)$ is the function which gives the absolute value of \(X\) 

- \(N\) is the number of forecasts. 

\vspace{5pt}

This way, the value of the loss function (MADL) is equal to the observed return on the investment with the predicted sign. This allows the model to inform us if its prediction will yield profit or loss and how much this profit or loss will be. MADL was designed specifically for working with AIS’s instead of just verification of forecasts in point. The function in our model is minimized, so that if it gives negative values, the strategy will make a profit, and if it gives positive values, the strategy will generate a loss.

\subsection{Ensemble models}
\label{Em}

In order to address the research questions we had to create two types of ensemble models: 

\begin{itemize}
\item type I - built with various theoretical models for the selected type of asset
\item type II - built with various types of assets for the selected theoretical model 
\end{itemize}

Therefore, type I ensemble models for a given asset $j$, were created according to the following formula:

\vspace{-10pt}
\begin{align}
\scriptsize
  \mathrm{EQline}^{\mathrm{(I)}}_j = \frac{1}{n} \sum_{i = 1}^{n}\mathrm{EQline}_{i,j}
\end{align}
\vspace{-10pt}

where:

$n$ - the number of theoretical models, $i = \{1, \dots, n\}$

$\mathrm{EQline}^{\mathrm{(II)}}_j$ - the value of the ensemble equity line on day $t$ for algorithmic investment strategy on the $j$-th asset (S\&P 500 index, Bitcoin, Gold, Natural Gas, and Wheat) for all theoretical models (LSTM, ARIMA-GARCH, Momentum, and Contrarian models),

$\mathrm{EQline}_{i,j}$ - the value of the single equity line on the day $t$ for algorithmic investment strategy on the $j$-th asset (S\&P 500 index, Bitcoin, Gold, Natural Gas, and Wheat) for the $i$-th theoretical model (LSTM, ARIMA-GARCH, Momentum, and Contrarian),

On the other hand, type II ensemble models for a given theoretical model $i$, were created according to the following formula:

\vspace{-10pt}
\begin{align}
\scriptsize
  \mathrm{EQline}^{\mathrm{(II)}}_i = \frac{1}{m} \sum_{j = 1}^{m}\mathrm{EQline}_{i,j}
\end{align}
\vspace{-10pt}

where:

$m$ - the number of assets, $j = \{1, \dots, m\}$

$\mathrm{EQline}^{\mathrm{(I)}}_i$ - the value of the ensemble equity line on day $t$ for algorithmic investment strategy on all assets (S\&P 500 index, Bitcoin, Gold, Natural Gas, and Wheat) for one of the $i$-th theoretical model (LSTM, ARIMA-GARCH, Momentum, and Contrarian). 

\subsection{Performance metrics}
\label{Pm}

\vspace{5pt}

Based on \cite{kijewskia2020predicting} the following performance metrics were calculated:

\begin{itemize}

    \item annualized return compounded (aRC):

\vspace{-10pt}
\begin{align}
\scriptsize
  \mathrm{aRC} = \prod_{i = 1}^{n}{(r_i+1)}^{252/n} - 1
\end{align}
\vspace{-10pt}

where $r_i$ is the daily percentage return at time $i$ and $n$ is the number of trading days,

    \item annualized standard deviation (aSD):

\vspace{-10pt}
\begin{align}
\scriptsize
  \mathrm{aSD} = \frac{\sqrt{252}}{n-1} \sum_{i = 1}^{n}{(r_i - \bar{r})^2}
\end{align}
\vspace{-10pt}

where $\bar{r}$ is the average daily percentage return,

    \item Information Ratio* (IR*):

\vspace{-10pt}
\begin{align}
\mathrm{IR}^{*} = \frac{\mathrm{aRC}}{\mathrm{aSD}}
\end{align}
\vspace{-10pt}

    \item Maximum Drawdown (MD):

\vspace{-10pt}
\begin{align}
\mathrm{MD} = \underset{0 \leq t_1 \leq t_2 \leq t}{sup}  \frac{EQ_{t_1} - EQ_{t_2}}{EQ_{t_1}}
\end{align}
\vspace{-10pt}

where $EQ_t$ is the equity line level at time $t$.



    \item Information Ratio** (IR**) 

\vspace{-10pt}
\begin{align}
\mathrm{IR}^{**} = \frac{\mathrm{aRC}*\mathrm{aRC}*\mathrm{sign}(\mathrm{ARC})}{\mathrm{aSD}*\mathrm{MD}}
\end{align}
\vspace{-10pt}

    We regard the IR** as the most important in the evaluation of our final results because this indicator combines the information from two crucial risk metrics: aSD and MD.

    \item Maximum Loss Duration (MLD): the longest time needed to surpass a maximum value of the strategy returns, measured in years.

    \item Information Ratio*** (IR***)

\vspace{-10pt}
\begin{align}
\mathrm{IR}^{***} = \frac{\mathrm{ARC}*\mathrm{ARC}*\mathrm{ARC}}{\mathrm{aSD}*\mathrm{MD}*\mathrm{MLD}}
\end{align}
\vspace{-10pt}

    \item nObs - the number of observation

    \item nTrades - the number of trades, which is the number of all changes in position on the analyzed asset

\end{itemize}

\subsection{Research description}
\label{Rd}

\vspace{3pt}

The detailed research description performed in this research can be summarized as follows:

\begin{itemize}
  \item tests for two versions of the investment strategies: Long Only and Short Only,
  \item a new Loss function: MADL, introduced by \cite{michankow2022etal}
  \item hyperparameters tuning, according to details described in Section \ref{FEHTaLt}
  \item walk-forward optimization:
  \begin{itemize}
      \item \textit{in-sample:} estimation of the model parameters (LSTM) or optimization of $p$ and $q$ orders (ARIMA-GARCH)
      \begin{itemize}
         \item in the \textit{in-sample} period we use the last $n \times 365$ actual days, where $n$ = {1, 2, 3, 4, 5}; base case = 3Y,
         \item in the \textit{in-sample} period we include data for the last 1Y, 2Y, 3Y, 4Y and 5Y years, respectively,
      \end{itemize}         
      \item \textit{out-of-sample:} re-estimation and re-optimization of models and forecast generations
      \begin{itemize}
        \item the first out-of-sample period starts one year after data start (for all five cases, i.e. 1Y, 2Y, 3Y, 4Y, and 5Y),
        \item out-of-sample forecasts: 1 day ahead,
      \end{itemize} 
  \end{itemize}  
  \item buy/sell signals definitions based on the next day forecasts,
  \item equity lines and performance metrics according to \cite{SlepaczukSakowskiZakrzewski+3918+36+55} with DFL =1,
  \item verification of diversification potential of various asset classes and theoretical models for AIS built for the S\&P500 index.
  \item the construction of ensemble investment strategies based on the combination of signals across different asset classes (S\&P500, BTC, UNG, GLD, ZWF) and theoretical models (ARIMA-GARCH, contrarian, momentum, and LSTM).
  \item the last part is devoted to the sensitivity analysis performed for various data frequencies used in the LSTM model.
\end{itemize}

\vspace{-15pt}





\section{Results}
\label{Results}

We present results from less complex (single investment models) to more complex (ensemble investment models) while emphasizing their diversification potential. This sequence is not necessarily connected with the order of our research questions in the Introduction.

\subsection{Base case scenario}
\label{Bcs}

In the first part of the results, we describe individual strategies and type I of the ensemble model where ensembling is defined as an equally weighted portfolio strategy, consisting of different models/strategies for a single asset. Rebalancing is performed on the first available day of Jan, Apr, Jul, and Oct. 

Table \ref{tab:base_case} shows the performance metrics for tested strategies (individual and ensemble - type I) and the benchmark Buy\&Hold strategy. Based on these results we can notice that the LSTM model-based strategy is characterized by the highest IR (IR*, IR**, and IR***) in most cases.

\begin{table}[H]
\caption{Base case scenario results for individual and ensemble strategies, for a single asset}
\centering
\begin{center}
\begin{tabular}[t]{lrrrrrrrrr}
\hline
  & \textbf{aRC} & \textbf{aSD} & \textbf{MD} & \textbf{MLD} & \textbf{IR*} & \textbf{IR**} & \textbf{IR***} & \textbf{nObs} & \textbf{nTrades}\\
\hline
\multicolumn{10}{l}{\textbf{BTC}}\\
\hspace{1em}B\&H & \textbf{114.78} & 88.29 & 86.67 & 3.24 & 1.30 & 1.722 & 0.610 & 3909 & 2\\
\hspace{1em}contra & -77.59 & 87.78 & 100.00 & 10.67 & -0.88 & -0.686 & -0.050 & 3909 & 1980\\
\hspace{1em}moment & 34.66 & 88.97 & 94.09 & 4.01 & 0.39 & 0.144 & 0.012 & 3909 & 1984\\
\hspace{1em}garch3 & -5.10 & 88.58 & 97.41 & 8.30 & -0.06 & -0.003 & 0.000 & 3909 & 1351\\
\hspace{1em}lstm & 109.32 & 64.76 & 67.19 & 2.89 & \textbf{1.69} & \textbf{2.747} & \textbf{1.040} & 3909 & 525\\
\hspace{1em}ensemble & 36.57 & 44.83 & 68.13 & 3.99 & 0.82 & 0.438 & 0.040 & 3909 & 6012\\
\multicolumn{10}{l}{\textbf{GLD}}\\
\hspace{1em}B\&H & \textbf{8.33} & 18.28 & 45.56 & 8.92 & \textbf{0.46} & \textbf{0.083} & 0.001 & 4117 & 2\\
\hspace{1em}contra & -9.68 & 17.74 & 87.46 & 14.05 & -0.55 & -0.060 & 0.000 & 4117 & 2133\\
\hspace{1em}moment & -17.83 & 18.85 & 96.20 & 16.21 & -0.95 & -0.175 & -0.002 & 4117 & 2143\\
\hspace{1em}garch3 & -14.29 & 18.28 & 92.88 & 16.27 & -0.78 & -0.120 & -0.001 & 4117 & 1593\\
\hspace{1em}lstm & 3.03 & 12.63 & 34.14 & 8.53 & 0.24 & 0.021 & 0.000 & 4117 & 720\\
\hspace{1em}ensemble & -8.95 & 6.10 & 78.95 & 16.27 & -1.47 & -0.166 & -0.001 & 4117 & 6853\\
\multicolumn{10}{l}{\textbf{SPX}}\\
\hspace{1em}B\&H & \textbf{10.35} & 19.48 & 55.25 & 4.48 & \textbf{0.53} & 0.100 & 0.002 & 4340 & 2\\
\hspace{1em}contra & -0.75 & 18.99 & 83.62 & 12.70 & -0.04 & 0.000 & 0.000 & 4340 & 2269\\
\hspace{1em}moment & -25.37 & 19.99 & 99.40 & 17.21 & -1.27 & -0.324 & -0.005 & 4340 & 2269\\
\hspace{1em}garch3 & 4.24 & 19.33 & 49.95 & 6.96 & 0.22 & 0.019 & 0.000 & 4340 & 1274\\
\hspace{1em}lstm & 7.23 & 14.92 & 28.43 & 1.99 & 0.48 & \textbf{0.123} & \textbf{0.004} & 4340 & 722\\
\hspace{1em}ensemble & -3.03 & 7.45 & 44.52 & 17.22 & -0.41 & -0.028 & 0.000 & 4340 & 6810\\
\multicolumn{10}{l}{\textbf{UNG}}\\
\hspace{1em}B\&H & -27.52 & 44.44 & 99.58 & 13.73 & -0.62 & -0.171 & -0.003 & 3512 & 2\\
\hspace{1em}contra & -5.47 & 43.86 & 94.69 & 9.64 & -0.12 & -0.007 & 0.000 & 3512 & 1803\\
\hspace{1em}moment & -30.88 & 45.05 & 99.70 & 13.93 & -0.69 & -0.212 & -0.005 & 3512 & 1797\\
\hspace{1em}garch3 & -19.90 & 44.40 & 97.75 & 9.61 & -0.45 & -0.091 & -0.002 & 3512 & 1416\\
\hspace{1em}lstm & \textbf{1.09} & 31.15 & 74.79 & 5.10 & \textbf{0.04} & \textbf{0.001} & 0.000 & 3512 & 612\\
\hspace{1em}ensemble & -8.05 & 14.98 & 79.28 & 9.61 & -0.54 & -0.055 & 0.000 & 3512 & 5852\\
\multicolumn{10}{l}{\textbf{ZWF}}\\
\hspace{1em}B\&H & \textbf{7.24} & 33.22 & 71.80 & 14.01 & \textbf{0.22} & \textbf{0.022} & 0.000 & 4362 & 2\\
\hspace{1em}contra & -14.61 & 32.65 & 95.09 & 16.94 & -0.45 & -0.069 & -0.001 & 4362 & 2240\\
\hspace{1em}moment & -21.97 & 33.82 & 99.17 & 13.99 & -0.65 & -0.144 & -0.002 & 4362 & 2264\\
\hspace{1em}garch3 & -29.25 & 33.21 & 99.78 & 17.18 & -0.88 & -0.258 & -0.004 & 4362 & 1818\\
\hspace{1em}lstm & 1.04 & 22.98 & 65.99 & 14.08 & 0.05 & 0.001 & 0.000 & 4362 & 778\\
\hspace{1em}ensemble & -12.68 & 10.61 & 90.69 & 17.18 & -1.20 & -0.167 & -0.001 & 4362 & 7376\\
\hline
\end{tabular}
\end{center}
\scriptsize
\justifying 
Note: Results cover the performance metrics for 4 individual strategies and 1 ensemble model for 5 various assets (BTC, GLD, SPX, UNG, and ZWF). The ensemble model stands for the combination of all theoretical models for the given asset.\label{tab:base_case}
\end{table}

\begin{figure}[H]
\centering
\includegraphics[width=\columnwidth]{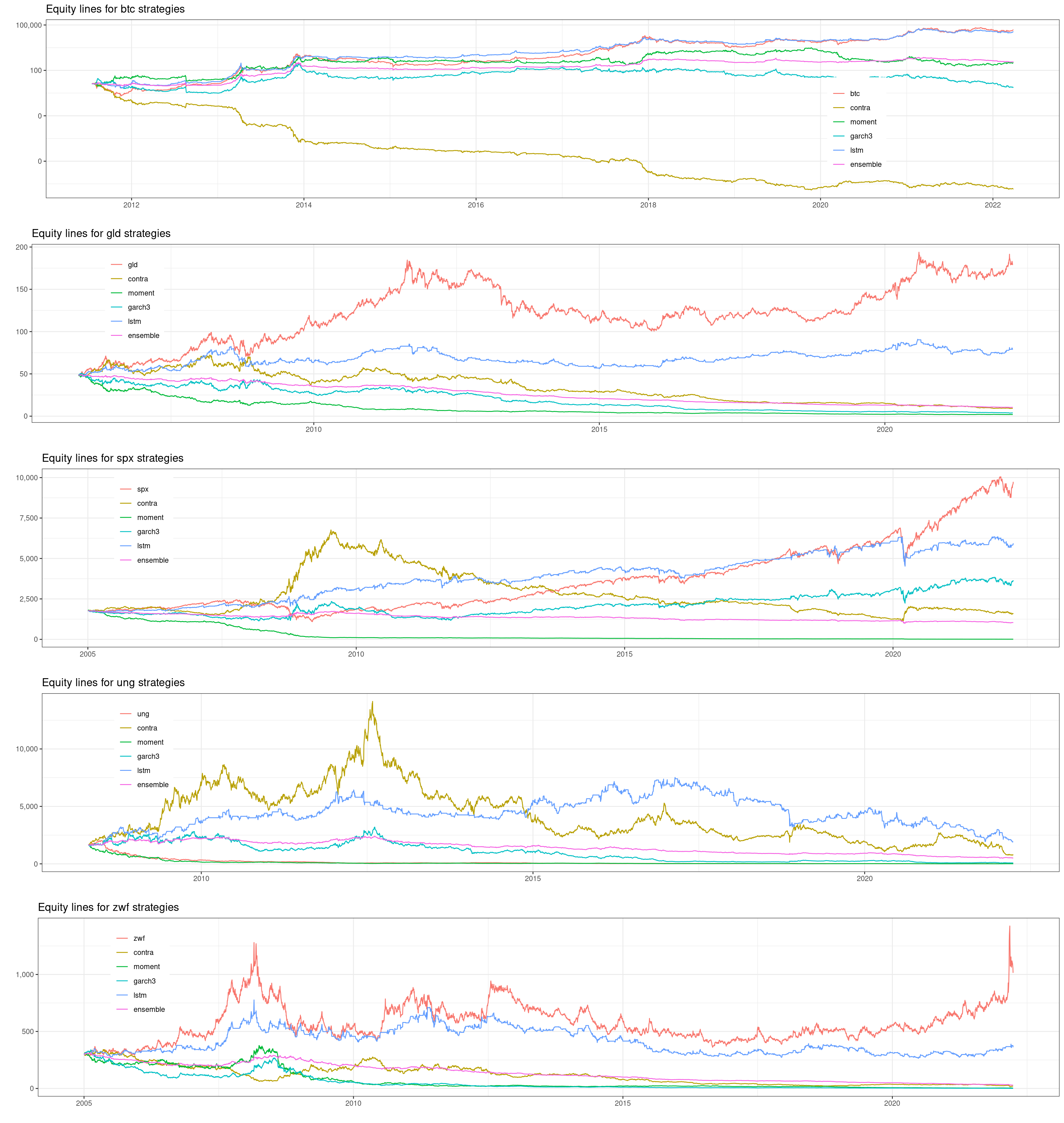}
\scriptsize
\justifying
Note: Each panel presents five equity lines for each tested asset (BTC, GLD, SPX, UNG, and ZWF). These equity lines represent the results for 4 individual strategies based on the model/concept of LSTM, ARIMA-GARCH, momentum, and contrarian, and one additional equity line for the ensemble model built using these four above-mentioned.
\footnotesize
\caption{Equity lines for individual and ensemble strategies for single assets}
\label{fig:EquityLinesBC}
\end{figure}

Figure \ref{fig:EquityLinesBC} presents equity lines for every investment strategy and confirms the results described in Table \ref{tab:base_case}.

Table \ref{tab:Ensemble_strategies} contains the performance metrics for all types of ensemble models (type I - ensemble model combining all tested strategies
for the given assets: SPX\_all, BTC\_all, GLD\_all, UNG\_all, ZWF\_all, and type II - ensemble model combining all assets for the given tested strategy: contr\_all, moment\-all, garch3\_all, lstm\_all
) and compare it with Buy\&Hold strategy for all 5 assets (5\_assets). The important conclusion from this table is that lstm\_all outperforms other strategies and Buy\&Hold and that BTC\_all outperforms other assets. The former could be attributed to the distinctive architecture of LSTM networks, which provides them with the capability to more effectively capture intricate temporal patterns within the data, while the latter to the availability and continuity of BTC data, which is quoted 24/7.

\begin{table}[H]
\caption{Ensemble strategies for single assets and theoretical models}
\centering
\begin{center}
\begin{tabular}[t]{lrrrrrrrrr}
\hline
  & \textbf{aRC} & \textbf{aSD} & \textbf{MD} & \textbf{MLD} & \textbf{IR*} & \textbf{IR**} & \textbf{IR***} & \textbf{nObs} & \textbf{nTrades}\\
\hline
\multicolumn{10}{l}{\textbf{B\&H all assets}}\\
\hspace{1em}B\&H\_all & \textbf{23.947} & 23.564 & 44.433 & 5.024 & 1.016 & 0.548 & 0.026 & 3908 & 10\\
\multicolumn{10}{l}{\textbf{ensembles for single assets}}\\
\hspace{1em}SPX\_all & -3.035 & 7.454 & 44.519 & 17.218 & -0.407 & -0.028 & 0.000 & 4340 & 6810\\
\hspace{1em}BTC\_all & 36.568 & 44.834 & 68.130 & 3.989 & 0.816 & 0.438 & 0.040 & 3909 & 6012\\
\hspace{1em}GLD\_all & -8.949 & 6.099 & 78.950 & 16.274 & -1.467 & -0.166 & -0.001 & 4117 & 6853\\
\hspace{1em}UNG\_all & -8.053 & 14.976 & 79.276 & 9.611 & -0.538 & -0.055 & 0.000 & 3512 & 5852\\
\hspace{1em}ZWF\_all & -12.684 & 10.608 & 90.690 & 17.179 & -1.196 & -0.167 & -0.001 & 4362 & 7376\\
\multicolumn{10}{l}{\textbf{models for all assets }}\\
\hspace{1em}contra\_all & -18.166 & 15.448 & 95.657 & 15.472 & -1.176 & -0.223 & -0.003 & 3908 & 10728\\
\hspace{1em}moment\_all & 2.567 & 22.708 & 68.705 & 5.980 & 0.113 & 0.004 & 0.000 & 3908 & 10760\\
\hspace{1em}garch3\_all & -2.862 & 23.202 & 83.842 & 12.028 & -0.123 & -0.004 & 0.000 & 3908 & 7755\\
\hspace{1em}lstm\_all & 19.674 & 18.072 & 30.274 & 4.460 & \textbf{1.089} & \textbf{0.707} & \textbf{0.031} & 3908 & 3660\\
\hline
\end{tabular}
\end{center}
\scriptsize
\justifying
Note: Each panel presents performance metrics for the Buy\&Hold strategy for all assets (5\_assets), for ensemble models for single assets (SPX\_all, BTC\_all, GLD\_all, UNG\_all, ZWF\_all), ensemble models for theoretical concepts (contra\_all, moment\_all, garch3\_all, lstm\_all).
\label{tab:Ensemble_strategies}
\end{table}

\begin{figure}[H]
\centering
\includegraphics[width=\columnwidth]{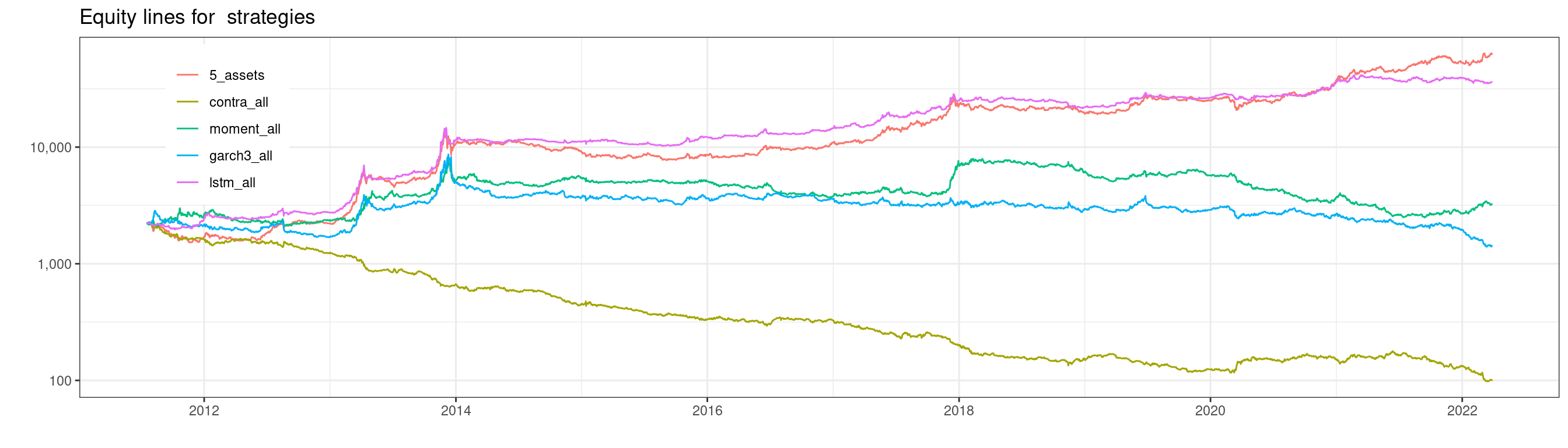}
\scriptsize
\justifying
Note: 5\_assets stands for Buy\&Hold strategy for all assets. Contra\_all, moment\_all, garch3\_all, lstm\_all stand for ensemble models for all assets within one theoretical concept. 
\footnotesize
\caption{Ensemble strategies for all theoretical models}
\label{fig:Ensemble_5_assets_all}
\end{figure}

Figure \ref{fig:Ensemble_5_assets_all} visualizes fluctuations of equity lines for ensemble models and Buy\&Hold and confirms the high performance of the LSTM model-based strategy.

Table \ref{tab:Ensemble_strategies_summary} contains a summary of the first part of the research which enables us to refer to our research questions.

\begin{table}[H]
\caption{Ensemble strategies for all assets within one theoretical model.}
\centering
\begin{center}
\begin{tabular}[t]{lrrrrrrrr}
\hline
& & & & & & \textbf{positive} & \textbf{beat} & \\
\textbf{IR**} & \textbf{BTC} & \textbf{GLD} & \textbf{SPX} & \textbf{UNG} & \textbf{ZWF} & \textbf{IR**} & \textbf{B\&H?} & \textbf{winner}\\
\hline
B\&H & 1.722 & \textbf{0.083} & 0.100 & -0.171 & 0.022 & 80\% & 0\% & 40\%\\
contrarian & -0.686 & -0.060 & 0.000 & -0.007 & -0.069 & 0\% & 20\% & 0\%\\
momentum & 0.144 & -0.175 & -0.324 & -0.212 & -0.144 & 20\% & 0\% & 0\%\\
garch3 & -0.003 & -0.120 & 0.019 & -0.091 & -0.258 & 20\% & 20\% & 0\%\\
lstm & \textbf{2.747} & 0.021 & \textbf{0.123} & \textbf{0.001} & {0.001} & {100\%} & {60\%} & {60\%}\\
ensemble & 0.438 & -0.166 & -0.028 & -0.055 & -0.167 & 20\% & 20\% & 0\%\\
\hline
\end{tabular}
\end{center}
\scriptsize
\justifying
Note: B\&H stands for Buy\&Hold strategy for all assets. Contrarian, momentum, garch3, lstm stand for ensemble models for all assets within one theoretical concept. The ensemble stands for ensemble model for all assets and all theoretical models.
\label{tab:Ensemble_strategies_summary}
\end{table}

Based on the results for the base case scenario, presented in Tables \ref{tab:base_case}, \ref{tab:Ensemble_strategies}, and \ref{tab:Ensemble_strategies_summary} and Figure \ref{fig:EquityLinesBC}, and \ref{fig:Ensemble_5_assets_all} we can refer to RQ2 and RQ3. 
Referring to RQ2, we can confirm that ML models are more efficient than classical models. In the case of single investment strategies because LSTM was the best strategy in 60\% of the cases (3 out of 5 asset classes tested). Moreover, in the case of complex investment strategies (type II) based on the aggregation of all assets for a single theoretical model lstm\_all was the best strategy in comparison to contrarian\_all, momentum\_all, and garch\_all.

Regarding RQ3, the ensemble AIS based on the aggregation of all theoretical models for the single asset (type I) were never better than the LSTM model or the B\&H strategy. Moreover, none of the ensemble strategies based on the aggregation of all assets for the single theoretical model (type II): lstm\_all, contrarian\_all, momentum\_all, garch\_all, and B\&H\_all were better than the single strategies for the given class of asset.

\subsection{Base Case Scenario. Ensemble models based on two assets - diversification potential.}
\label{DP}

\begin{figure}[H]
\centering
\includegraphics[width=\columnwidth]{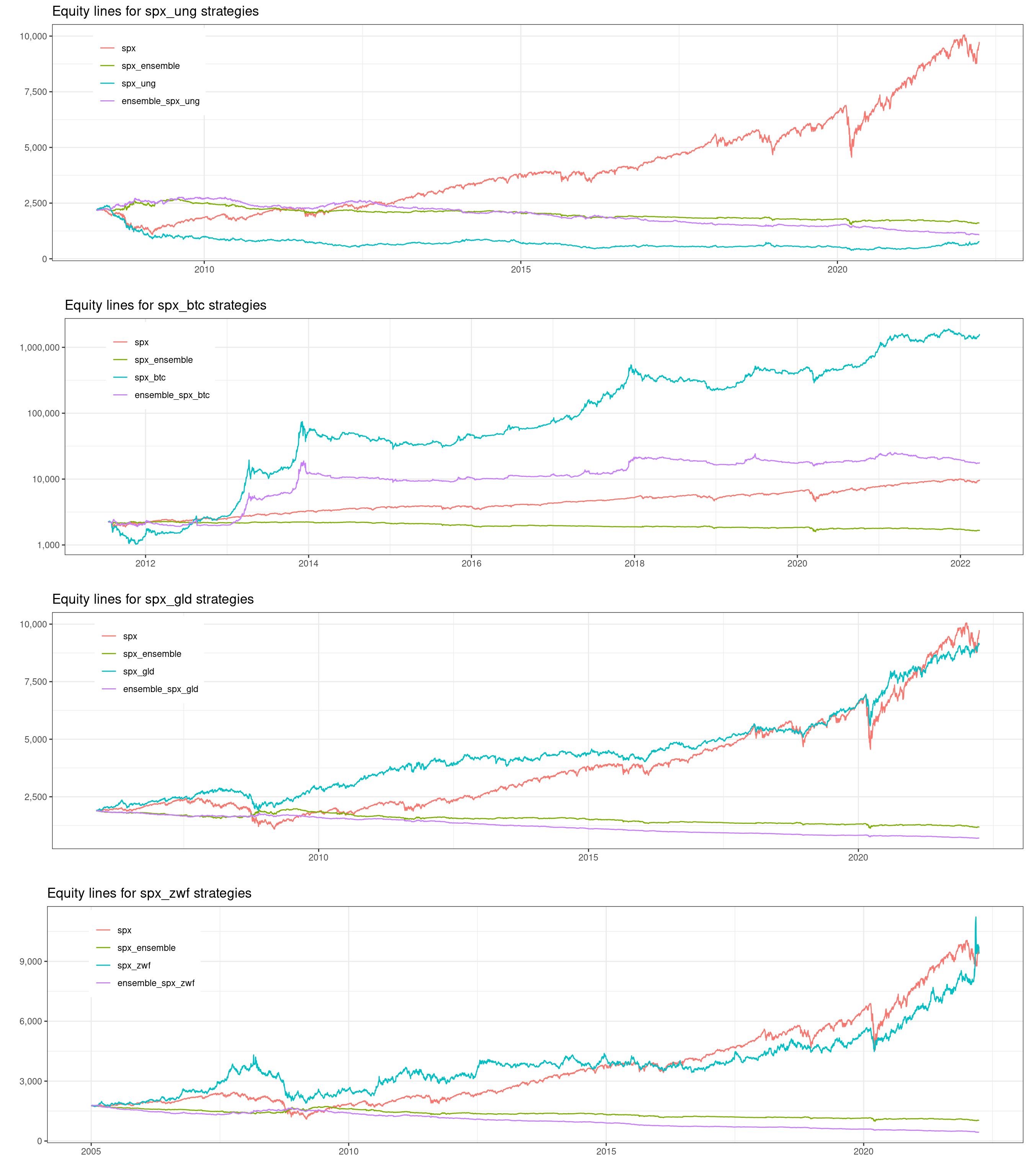}
\scriptsize
\justifying
Note: Each panel presents the equity lines for 4 different strategies: SPX - B\&H for the S\&P 500 index, spx\_ensemble - the ensemble models combining all theoretical models for the S\&P 500 index, spx\_asset - combined B\&H for S\&P 500 index and the given asset, ensemble\_spx\_asset - the combination of two ensemble models built for all theoretical models for SPX and the given asset.
\footnotesize
\caption{Equity lines for hedging strategies for equity index}
\label{fig:diversification_potential_fig}
\end{figure}

Based on the results presented in Table \ref{tab:diversification_potential} and Figure \ref{fig:diversification_potential_fig} for the ensemble models of two assets and their diversification potential with regard to strategies based on SPX, we can refer to the RQ1. Looking at the IR** measure, we can state that the only diversification potential can be noticed after adding the ensemble model based on BTC to the ensemble model based on SPX, where the IR** for the ensemble\_spx\_btc increases.

\begin{table}[H]
\caption{Diversification potential of investment models for hedging equity index investment model}
\centering
\begin{center}
\begin{tabular}[t]{lrrrrrrrrr}
\hline
  & \textbf{aRC} & \textbf{aSD} & \textbf{MD} & \textbf{MLD} & \textbf{IR*} & \textbf{IR**} & \textbf{IR***} & \textbf{nObs} & \textbf{nTrades}\\
\hline
\multicolumn{10}{l}{\textbf{UNG}}\\
\hspace{1em}spx & \textbf{11.27} & 20.59 & 51.52 & 2.75 & \textbf{0.55} & \textbf{0.120} & 0.005 & 3512 & 2\\
\hspace{1em}spx\_ensemble & -2.14 & 7.90 & 42.82 & 12.69 & -0.27 & -0.014 & 0.000 & 3512 & 5357\\
\hspace{1em}spx\_ung & -7.39 & 25.30 & 84.11 & 13.77 & -0.29 & -0.026 & 0.000 & 3512 & 224\\
\hspace{1em}ensemble\_spx\_ung & -4.91 & 8.25 & 61.09 & 12.59 & -0.59 & -0.048 & 0.000 & 3512 & 11209\\
\multicolumn{10}{l}{\textbf{BTC}}\\
\hspace{1em}spx & 10.01 & 14.44 & 33.79 & 1.08 & 0.69 & 0.205 & 0.019 & 3908 & 2\\
\hspace{1em}spx\_ensemble & -1.76 & 5.64 & 30.19 & 14.37 & -0.31 & -0.018 & 0.000 & 3908 & 4003\\
\hspace{1em}spx\_btc & \textbf{52.72} & 42.33 & 61.87 & 4.42 & \textbf{1.25} & \textbf{1.061} & \textbf{0.127} & 3908 & 172\\
\hspace{1em}ensemble\_spx\_btc & 14.36 & 23.01 & 52.71 & 5.82 & 0.62 & 0.170 & 0.004 & 3908 & 10012\\
\multicolumn{10}{l}{\textbf{GLD}}\\
\hspace{1em}spx & \textbf{10.58} & 19.85 & 55.25 & 4.48 & 0.53 & 0.102 & 0.002 & 4117 & 2\\
\hspace{1em}spx\_ensemble & -2.72 & 7.61 & 42.82 & 12.69 & -0.36 & -0.023 & 0.000 & 4117 & 6417\\
\hspace{1em}spx\_gld & 10.17 & 13.58 & 34.05 & 1.72 & \textbf{0.75} & \textbf{0.224} & \textbf{0.013} & 4117 & 264\\
\hspace{1em}ensemble\_spx\_gld & -5.80 & 5.00 & 63.21 & 16.31 & -1.16 & -0.107 & 0.000 & 4117 & 13270\\
\multicolumn{10}{l}{\textbf{ZWF}}\\
\hspace{1em}spx & \textbf{10.30} & 19.43 & 55.25 & 4.48 & \textbf{0.53} & \textbf{0.099} & \textbf{0.002} & 4362 & 2\\
\hspace{1em}spx\_ensemble & -3.02 & 7.44 & 44.52 & 17.31 & -0.41 & -0.028 & 0.000 & 4362 & 6802\\
\hspace{1em}spx\_zwf & 10.06 & 20.45 & 56.04 & 6.81 & 0.49 & 0.088 & 0.001 & 4362 & 276\\
\hspace{1em}ensemble\_spx\_zwf & -7.80 & 6.54 & 75.55 & 17.18 & -1.19 & -0.123 & -0.001 & 4362 & 14178\\
\hline
\end{tabular}
\end{center}
\scriptsize
\justifying
Note: Each of the 4 panels contains the results for 4 strategies: SPX - B\&H for S\&P 500~index, spx\_ensemble - the ensemble models combining all theoretical models for S\&P 500 index, spx\_asset - combined B\&H for S\&P 500 index and the given asset, ensemble\_spx\_asset - the combination of two ensemble models built for all theoretical models for SPX and the given asset. 
\label{tab:diversification_potential}
\end{table}

\subsection{Daily versus hourly results for selected assets}
\label{DvH}

In order to answer RQ4, we repeat training and estimation of the LSTM model for SPX and UNG assets on hourly data in the same period as for the daily data, i.e. from 2008-04-17 to 2022-03-29. The selection of these two assets was dictated by the following reasons. The S\&P 500 index was chosen for its wide usage in financial literature, ensuring comparability with other research. UNG represents a distinct dynamic with decreasing asset prices over time and potential diversification benefits during geopolitical stress, such as the Russian-Ukrainian conflict.

\begin{table}[H]
\caption{LSTM model results for S\&P 500 index and UNG on daily and hourly data}
\begin{center}
\begin{tabular}{lrrrrrrrrr}
\hline
                 & \textbf{aRC} & \textbf{aSD} & \textbf{MD} & \textbf{MLD} & \textbf{IR*} & \textbf{IR**} & \textbf{IR***} & \textbf{nObs} & \textbf{nTrades} \\ \hline
\multicolumn{10}{l}{\textbf{S\&P 500}}\\
\hspace{1em}lstm\_1d         & 7.23         & 14.92        & 28.43        & 1.99        & 0.48         & 0.123         & 0.004          & 4340          & 722              \\
\hspace{1em}lstm\_1h         & \textbf{9.72}         & 12.34        & 24.25        & 1.72        & \textbf{0.79}         & \textbf{0.315}         & \textbf{0.018}          & 34702         & 5364             \\
\multicolumn{10}{l}{\textbf{UNG}}\\
\hspace{1em}lstm\_1d         & 1.09         & 31.15        & 74.79        & 5.1         & 0.04         & 0.001         & 0.000          & 3512          & 612              \\
\hspace{1em}lstm\_1h         & \textbf{6.80}          & 24.38        & 70.54        & 6.17        & \textbf{0.28}         & \textbf{0.027}         & 0.000          & 122318        & 16310            \\ \hline
\end{tabular}
\end{center}
Note: lstm\_1d stands for LSTM model-based investment strategy trained and estimated for the S\&P 500 index (first panel) and UNG on daily data. lstm\_1h denotes the same strategy test on 1h data.
\label{tab:LST_1h_tab}
\end{table}

\begin{figure}[H]
\centering
\includegraphics[width=\columnwidth]{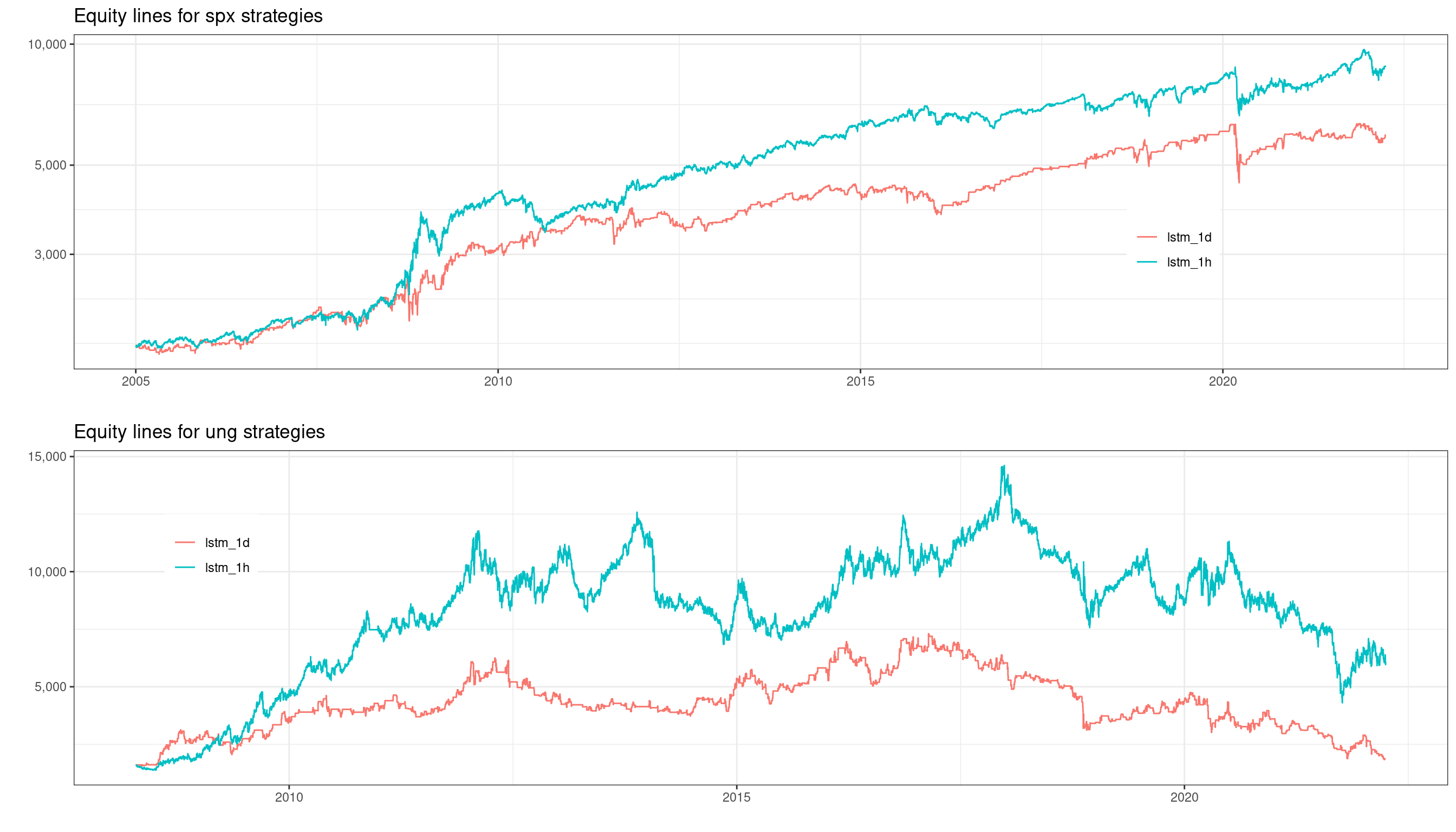}
\scriptsize
\justifying
Note: Each panel presents equity lines for SPX and UNG for two different frequencies daily and hourly.
\footnotesize
\caption{Equity lines for LSTM model results for S\&P 500 index and UNG on daily versus hourly data}
\label{fig:LSTM_1h_fig}
\end{figure}

Table \ref{tab:LST_1h_tab} and Figure \ref{fig:LSTM_1h_fig} shows that in each tested case LSTM models on hourly data outperform the ones on daily data in each case of risk-adjusted measures (IR*, IR**, and IR***).

\vspace{-10pt}

\section{Conclusions}
\label{Conclusions}

The novelty and the main contribution of this paper is an attempt to focus on the problem of diversification from a different perspective than what is usually presented in state-of-the-art research. Based on the results for five different assets (BTC, GLD, SPX, UNG, and ZWF), in the period from 2007 to 2022, we verified a few different research questions focusing on individual and ensemble algorithmic investment strategies using various types of theoretical models. The ensemble process used in this research for the first time focused on 3 different surfaces of single strategies combination, i.e. based on 1) various types of assets, 2) various theoretical models, and 3) a combination of both of them.

We verify the diversification potential of investment strategies for the equity index (S\&P 500 index) based on various theoretical concepts against other investment strategies (RQ1). Therefore, referring to RQ1: \textit{Which of the tested groups of assets (energy commodities, cryptocurrencies, gold, or soft commodities) have the largest diversification potential in the complex algorithmic investment strategies, built with machine learning models and ARIMA-GARCH models for equity indices?}, based on the results presented in Table \ref{tab:diversification_potential} and Figure \ref{fig:diversification_potential_fig}, we can state that only ensemble\_BTC has the diversification potential that increases the efficiency of ensemble models for the equity index. Moreover, taking into account that the distribution of returns for other equity indices is quite similar to that of the S\&P 500 we are sure that our conclusions can be extended to them, as well.

Based on the results presented in Table \ref{tab:Ensemble_strategies_summary} and Figure \ref{fig:Ensemble_5_assets_all} we can affirmatively address RQ2: \textit{Are machine learning techniques more efficient than ARIMA-GARCH models and the concepts of momentum and contrarian in the case of single and complex (ensemble model combining all tested strategies for the given assets - type I) investment strategies}

After analyzing the results presented in Table \ref{tab:base_case} and Table \ref{tab:Ensemble_strategies_summary} we can assert an unfavorable response to RQ3: \textit{Are complex (ensemble) AIS based on the aggregation of all theoretical models for the single asset (type I) or all assets for a single theoretical model (type II) more efficient than individual strategies.}

Finally, based on the results presented in (Table \ref{tab:diversification_potential} and Figure\ref{fig:diversification_potential_fig}) we can provide a positive response to RQ4: \textit{Are results for LSTM models on higher frequencies of data (1h) better than those on daily data.}

Further research extensions of this work should focus on the following: extensive sensitivity analysis with a special focus on alternative loss functions, a larger set of alternative assets and models, different theoretical models in the process of generating buy/sell signals, more careful hyperparameters tuning process, and finally more advanced procedure of selection parameters and hyperparameters in the in-sample period.

\bibliography{bibfile}
\end{document}